\documentclass[preprint,12pt]{elsarticle}

\usepackage{amssymb}
\journal{High Energy Density Physics }

\begin{document}

\begin{frontmatter}

\title{On plasma radiative properties in stellar  conditions}

\author{S. Turck-Chi\`eze\corref{cor1}$^1$, F. Delahaye $^1$, D. Gilles $^1$, G. Loisel $^{1,2}$, L. Piau$^1$}
\cortext[cor1]{Corresponding author  E-mail address:  sylvaine.turck-chieze@cea.fr \\Tel.: 33 1 69 08 43 87}
\address{\rm CEA/DSM/IRFU $^1$ and IRAMIS $^2$, CE Saclay 91191 Gif sur Yvette, France\fnref{label1}}
\begin{abstract}
The knowledge of stellar evolution is evolving quickly thanks  to an increased number of opportunities to scrutinize the stellar internal plasma properties by stellar seismology and by 1D and 3D simulations.  These new tools help us to introduce the internal dynamical phenomena in stellar modeling. A proper inclusion of these processes supposes a real confidence in the microscopic physics used, partly checked by  solar or stellar acoustic modes.  In the present paper we first recall  which fundamental physics has been recently verified by helioseismology.  Then we recall  that opacity is an important ingredient of the secular evolution of stars and we point out why it is necessary to measure absorption coefficients and degrees of ionization in the laboratory for some well identified astrophysical conditions. We examine two specific experimental conditions which are accessible to large laser facilities and are suitable to solve some interesting  questions of the stellar  community: 	are the  solar internal radiative interactions properly estimated and what is the proper  role of the opacity in the excitation of the non radial modes in the envelop of the $\beta$ Cepheids  and the  Be stars ? At the end of the paper we point out the difficulties of the experimental approach that we need to overcome. \end{abstract}


\end{frontmatter}

\section{The context}
\label{context}
The stellar structure is currently determined by the resolution of  four  equations which govern the 
hydrostatic equilibrium, the conservation of mass, the energetic equilibrium and the energy transport dominated by radiation or convection. In order to describe the secular temporal evolution of stars, one needs to add to these coupled equations the temporal evolution of the composition through nuclear reaction rates and diffusion or turbulent effects  \cite{TC93}. In fact the resolution of  these equations supposes a good description of plasma properties on a very broad range of  temperature (typically several thousand to billion degrees) and density (10$^{-7}$ to 1000 g/cm$^3$) in term of nuclear interaction, equation of state and photon interaction with the different elements of the stellar plasma.
Today  helio and astero seismology allow to check  the hypotheses of such a calculation and new indicators  are extracted from helioseismology to progress towards a more dynamical view of stars  \cite{Vorontsov02, Mathur08}. Such approach is now extended to other stars with a better understanding of massive stars \cite{Townsend07,Ud-Doula08}. This crucial effort gives us the potential to solve new problems in our knowledge of stars:

\noindent
- the development of a unified theory of stellar evolution will naturally link  short time stages dominated by the dynamics (stellar formation, young stars modeling or supernovae explosion) and long stages dominated by slow microscopic evolution but where dynamical effects continue to be present,

\noindent
- the complete MHD description of the Sun will connect internal and external dynamics. This will help to establish the origin(s) of the solar activity and its prediction for the next century with its consequent impact on the terrestrial environment. Such approach will also be useful for understanding stellar winds in massive stars which turn quickly, are deformed and are magnetically active.

Consequently, we are currently in a period of  transition where we evolve from a qualitative description of stars to an entirely quantitative one.  On one hand, one can say that
some plasma properties are definitively established (section 2).  On the other hand, internal rotation and magnetic field induce transport of momentum and chemicals which are still difficult to estimate properly due to our poor knowledge of the initial conditions and of the increased number of transport coefficients in the equations. 
So there is a real benefit to isolate some phenomena and go to the laboratory for a verification or an improvement of the theoretical predictions. 

Three decades ago,  stellar evolution drove measurements of  nuclear cross sections in the laboratory. This activity is now oriented toward the ultimate stages of evolution.  Today large laser facilities have become very crucial tools to check the absorption coefficients in some specific conditions, this information is fundamental to verifying the transport of energy inside the stars by photon interaction and to improve our description of the stellar envelops.
Section 3 will examine how the absorption coefficients are introduced in the stellar equations, we deduce from them two examples where opacity coefficients must be controlled. The astrophysical need leads to a campaign of absorption coefficient measurements as fundamental checks of atomic physics in stellar plasma conditions like it has been done for nuclear cross sections in the last two decades.

Section 4 will focus on the specific needs and on the natural limitations of the experimental side that we must reduce if we want that these experiments will contribute, among others, to  the renewal of stellar evolution.

\section{Stellar plasma properties checked by helioseismology}
Helio(astero) seismology is a discipline exploring the standing waves which are permanently excited by the superficial or deep convection of the Sun or stars. It  has revealed interesting internal properties of the solar plasma.

Acoustic waves propagate
inside the whole Sun and their travel time depends on the local sound speed. 
They generate very slightly motions in the atmosphere of
 the Sun which are detectable. 
 We can formely treat this information through a perturbative theory because these perturbations are small and the Sun is reasonably spherical.

The Sun (or  a solar like star) is a self-graviting sphere of compressible gas; it oscillates around its 
equilibrium state with a period of about 5-min. These oscillations are 
interpreted as a superposition of waves propagating inside the star (acting as a 
resonant cavity), and forming standing waves: the eigenmodes of vibration. By 
projecting these modes onto spherical harmonics $Y_l^m$, we write any scalar 
perturbations as (in the case of the pressure $p^{'}$) \cite{CDB91}:
\[p^{'}(r,\theta,\varphi,t)=p^{'}(r)Y_l^m(\theta,\varphi)\exp i\omega_{n,l,m} t \]
and the vector displacement $\vec{\xi}$ as
$$\vec{\xi}(r,\theta,\varphi,t)=\left(\xi_r(r),\xi_h(r)\frac{\partial}{\partial 
\theta},\xi_h(r)\frac{\partial}{\sin \theta \partial 
\varphi}\right)Y_l^m(\theta,\varphi)\exp i\omega_{n,l,m} t \quad  \eqno(1) $$
where $\xi_h=1/(\omega^2 r)[p^{'}/\rho+ \Phi^{'}]$ is the horizontal 
displacement, $\Phi^{'}$ the gravitational potential perturbation, 
$\omega_{n,l,m}$ the eigenfrequency and $\rho$ the gas density. The quantum 
numbers $n$, $l$, $m$ are respectively the radial order (number of nodes along 
the radius), the degree (the total horizontal wave number at the 
surface is $k_h\sim L/R_{\odot}$, with  $L=\sqrt{l(l+1)}$) and the azimuthal 
order (number of nodes along the equator with $|\vec{m}| \le l$). Restricting the phenomenon to adiabatic 
oscillations within the Cowling approximation ($\Phi^{'}$ neglected) and 
considering only small radial wavelengths compared to $R_{\odot}$, we reduce a 4th-order system equations to a 
second-order wave equations, with the following 
dispersion relation \cite{CDB91}
$$k^2_r=\frac{1}{c^2_s}\left[F^2_l\left(\frac{N^2}{\omega_{n,l,m}^2}-1\right)+\omega_{n,l,m}^2-\omega^2_c\right] \quad \eqno(2) $$
where the squared length of the wave vector is written as the sum of a radial and 
a horizontal component: $|\vec{k}|=k^2_r+k^2_h$, $k^2_h=F^2_l/c^2_s$ is the 
horizontal wave number, $F^2_l=L^2c^2_s/r^2$ the Lamb frequency, 
$N^2=g[1/\Gamma_1 d\ln p/dr - d\ln\rho/dr]$ the Brunt-V\"ais\"al\"a frequency, 
$\omega^2_c=c^2_s(1-2dH_{\rho}/dr)/4H^2_{\rho}$ the acoustic cut-off frequency ($\sim 
5.8$ mHz), $H^{-1}_{\rho}=-d\ln \rho/dr$ the density scale height, $\Gamma_1$ 
the adiabatic exponent and $c^2_s=\Gamma_1 p/\rho$ the sound speed.
\begin{figure}
\vspace{-0.5cm}
\begin{center}
\setlength{\unitlength}{1.0cm}
\hspace{-1.2cm}  
\includegraphics[width=6cm, angle=90]{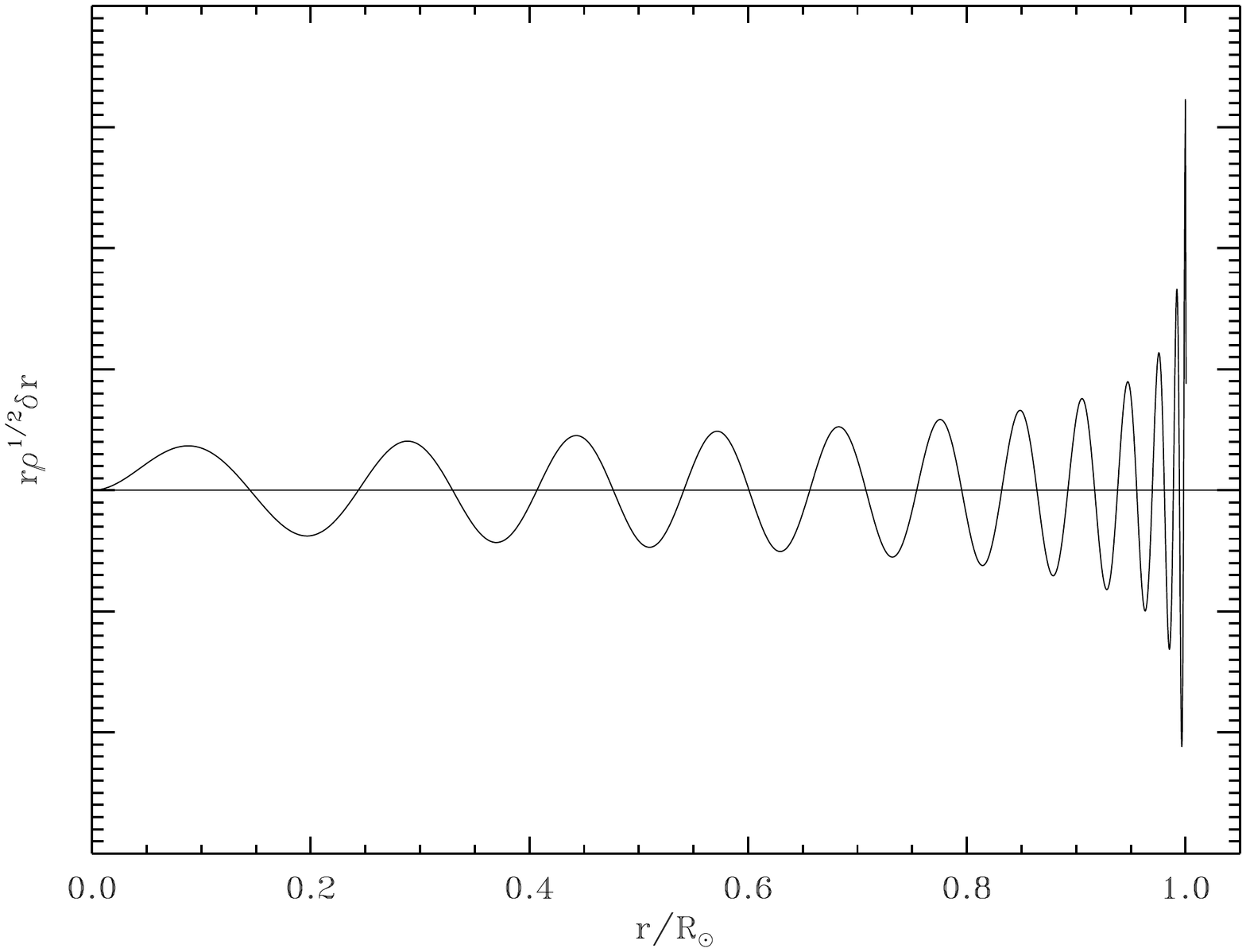} \hspace{-1.4cm}  
 \includegraphics[width=6.cm, angle=90]{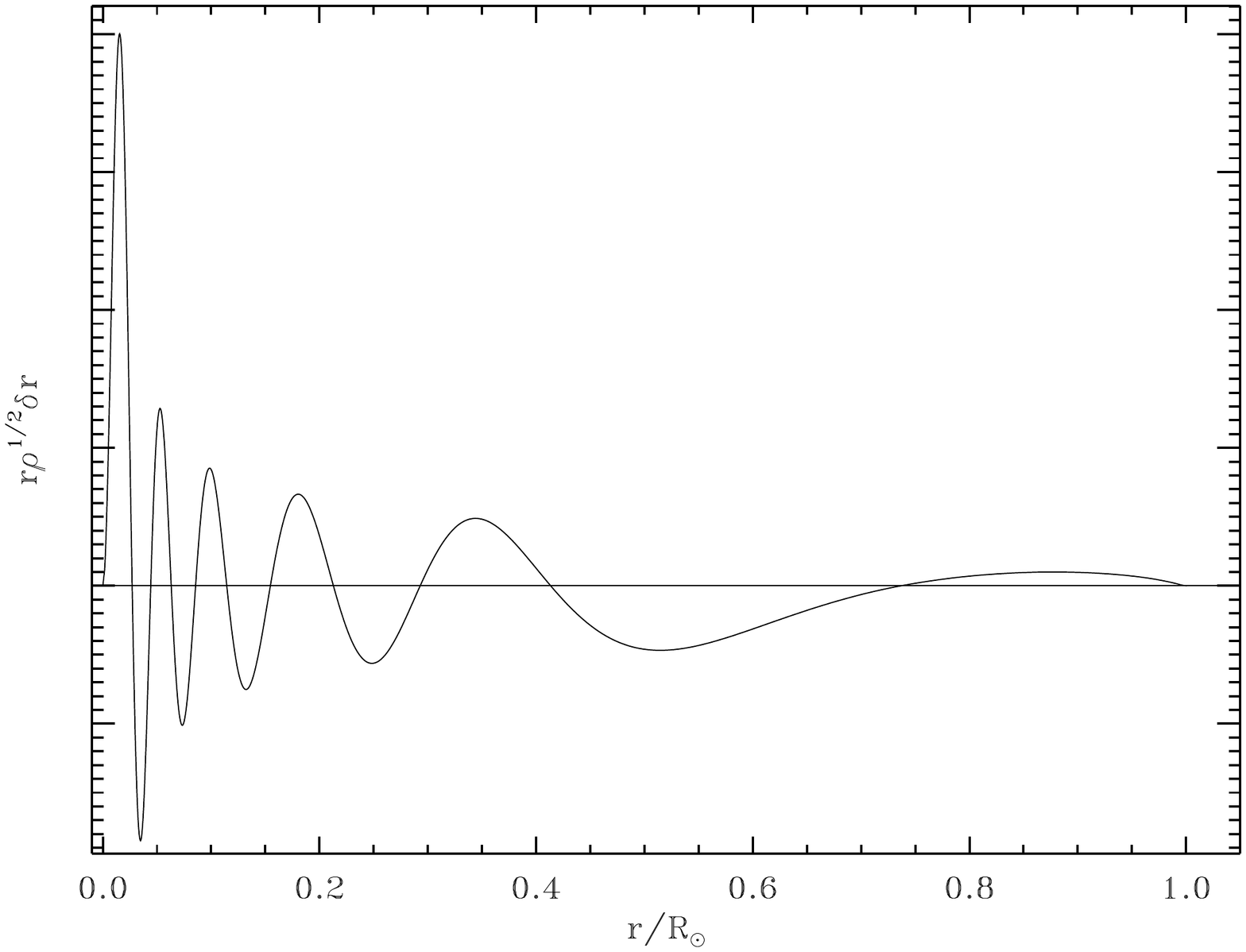}

\vspace{-0.5cm}
 \caption{Eigenfunctions of an acoustic mode ($\ell$ =0 radial mode, n=23), and of 
 a gravity mode ($\ell$ =2, n=10). These modes are now detected in different stellar sites.}
  \label{fig:modes}
\end{center}
 \vspace{-1.cm} 
\end{figure}
The oscillatory solutions of the wave equation define two types of waves, i.e 
acoustic ones (with $\omega_{n,,\ell,m}> N, F_\ell$) and gravity ones (with $\omega_{n,\ell,m}< N, F_\ell$).

Figure 1 shows that the acoustic modes have their maximum amplitude at the surface
(left side) but the low $\ell$ modes penetrate down to the core. The gravity modes are excellent probes of the solar (stellar) core and are evanescent at the surface
with rather small amplitude (right side).
A lot of solar acoustic modes  and some gravity modes have been detected. So a refined analysis of their properties leads to a 
stratified analysis of the solar internal structure from the surface down to the inner 6\% of the radius thanks to a very high accuracy (better than 10$^{-4}$) extraction of the sound speed  \cite{TC01a}. Figure 35 of \cite{TC93} shows the sensitivity of  the sound speed profile through the ratio $T/ \mu$ where $\mu$ is the mean molecular weight to physical processes. The inflexion of the profile due to the partial ionisation of helium  or due to the change of energy transport between radiation and convection (see Figure 46 of \cite{TC93}) has led to the determination of the solar helium content and of the position of the basis of the convective zone. Recently, two other important properties of the dense central plasma have been checked from the sound speed profile of the nuclear core: the maxwellian distribution of the reactant particle velocities and the absolute value of the proton proton reaction rate \cite{TC01b} in order to properly predict the solar neutrinos and also the central temperature.
\begin{figure}
\begin{center}
\includegraphics[width=7.cm]{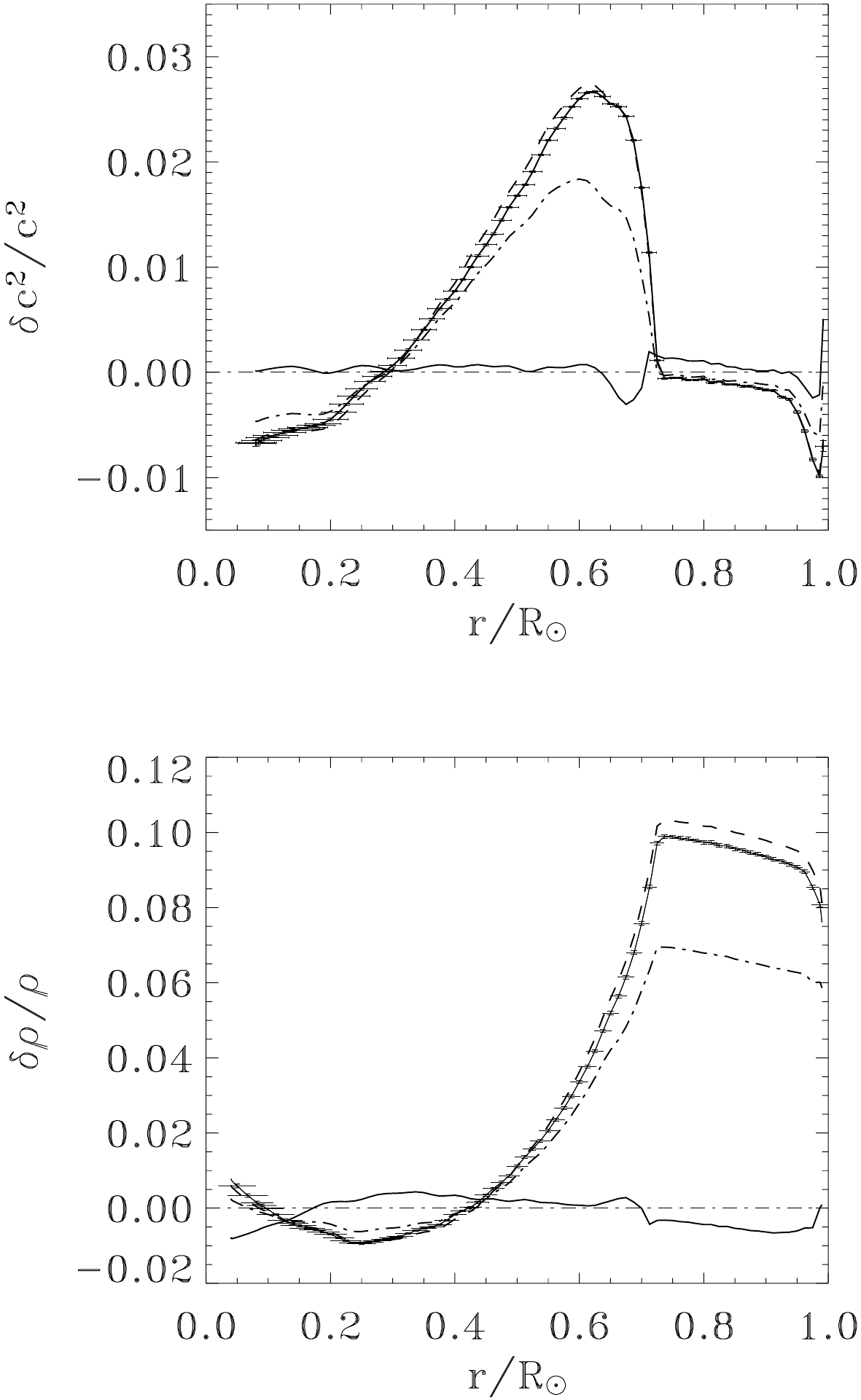}
\caption{Squared sound speed  and density difference deduced from the SoHO observations and the solar model using 2 updated compositions: Holweger (- . - .-) and Anders (full line + error bars) or with the seismic model. See details in  \cite{TC04}. The recent reanalysis of the solar oxygen and nitrogen estimate favors a difference similar to the one called here Holweger.}
\end{center}
\end{figure}

The sensitivity of the sound speed to the different ingredients of the solar model has been studied in \cite{TC01}.  This observable is particularly sensitive to the detailed solar composition through the opacity coefficients and one shows that a 1\%variation of the opacity coefficient produces an 0.1\% variation of the sound speed which is known with an accuracy of some 10$^{-5}$ \cite{TC01a}. The use of the updated determination of  the CNO superficial composition has led to a large discrepancy between the observed sound speed and the sound speed extracted from the standard solar model. In fact this discrepancy is  about 15 to 26 times the vertical uncertainty as shown on Figure 2 \cite{TC04}. The two   recent estimates of the solar composition are now confirmed by a third estimate and the discrepancy converges towards the curve corresponding to what we call here the Holweger composition. 

The interpretation of this discrepancy is not yet understood. A lot of potential interpretations have been studied, see for example \cite{Delahaye} and the review of \cite{Turck08} which gives a summary of the present hypotheses. The origin could be an insufficient knowledge of the composition inside the radiative zone or of the related opacity coefficients. The effect is  about 20-30\% or 10\% on the mean Rosseland values depending if the problem is connected to only one element or to most of the heavy elements (Z $>$ 2). Another explanation could be that the solar model  is incomplete because the dynamical effects are not yet introduced and could provide some transport of the different species. Helioseismology cannot solve this problem due to the connection between abundance and opacity within the solar model. Consequently this important discrepancy encourages us to study some experimental conditions in the laboratory and to measure  appropriate absorption spectra to validate or improve the complex theoretical estimates.

\section{The stellar opacities}
Stellar opacities partly define the longevity of stars. In fact the lifetime of the stars depends on the stellar mass and   the heavy element contribution to the total opacity coefficient while 
the stellar composition is generally dominated by hydrogen and helium. In fact, except below the surface, the light elements are generally totally ionized so  their contribution to the absorption coefficients is easier to estimate.
Inside stars, the energy transfer in the radiative zones  follows the equation:
$$ {dT\over dr} ={
{{-3}\over {4 ac}} {{\kappa_R \rho }\over {T^{3}}} {L(r)\over{4 \pi r^2}}}  \rm \;where \; {1\over \kappa_R} = {{\int_0^\infty  {1 \over {\kappa_{\nu a}[ 1-exp(-h \nu/kT)] +\kappa_{\nu s}}}  {dB_\nu \over dT} d\nu} \over {\int_0^\infty  {dB_\nu \over dT} d \nu} } \; \eqno{(3)}$$
 $\kappa_R$ is the Rosseland opacity at the position r and for a mixture of  species X$_i$. The weighting factor $dB_\nu \over dT$   gives the greatest weight to the region around  $\nu= 4 kT/h$.  L, T, $\rho$  are the local luminosity, temperature and density of the plasma at the position r. In these conditions, a species which is not totally ionized and has a large Z (number of protons) value may play an important role \cite{Courtaud}. $\kappa_{\nu a}$ is the term of absorption, it includes free-free, bound-free and bound-bound processes and $\kappa_{\nu s}$ is the scattering term.

\subsection{Sun and solar-like stars} 
 The calculation of the opacity coefficients is complex for solar like stars because the plasma is generally partially ionized so the astrophysical community has greatly benefitted from the work of the Los Alamos team (LAOL tables) \cite{Cox} and more recently from the OPAL team \cite{Iglesias, Rogers} who has delivered tables for a large range of temperature, density and composition since 1996. A third academic team  \cite{Seaton}  has first worked on the opacities of the stellar envelop and extended their work to  all stellar conditions in providing the OP tables. This team provides also the corresponding spectra \cite{Badnell}.
 
 Figure 3 shows the respective role of different elements in the internal solar conditions for the two used  solar composition and OPAL opacities \cite{TC04, TC05, TC08}, the same estimate using the OP opacities leads to slight differences mainly for the oxygen contribution which varies by a little more than 10\% with an impact of the total Rosseland coefficient of about 4\% near the convective basis \cite{Delahaye05}.  The large contribution of iron, neon and oxygen in the radiative zone (below 0.71 R$_\odot$) is due to their bound-bound contributions. Silicon also plays a  role at a level of 10\% below $10^{7°}$K. It is interesting to note that  iron has a fraction number of $5 \,10^{-5} $  and is always partly ionized even in the center of the Sun, its contribution to the total opacity is at a level of 20\% in most of the radiative zone. Oxygen becomes partly ionized at 0.4 R$_\odot$ and plays the major role to limit the radiative transfer, around 0.71 R$_\odot$. The increase of its opacity contribution  triggers the convection which is the dominant process in the transport of energy in the outer layers.
\begin{figure}
\vspace{-0.5cm} \hspace{-0.8cm} \includegraphics[ width=6cm, angle=90]{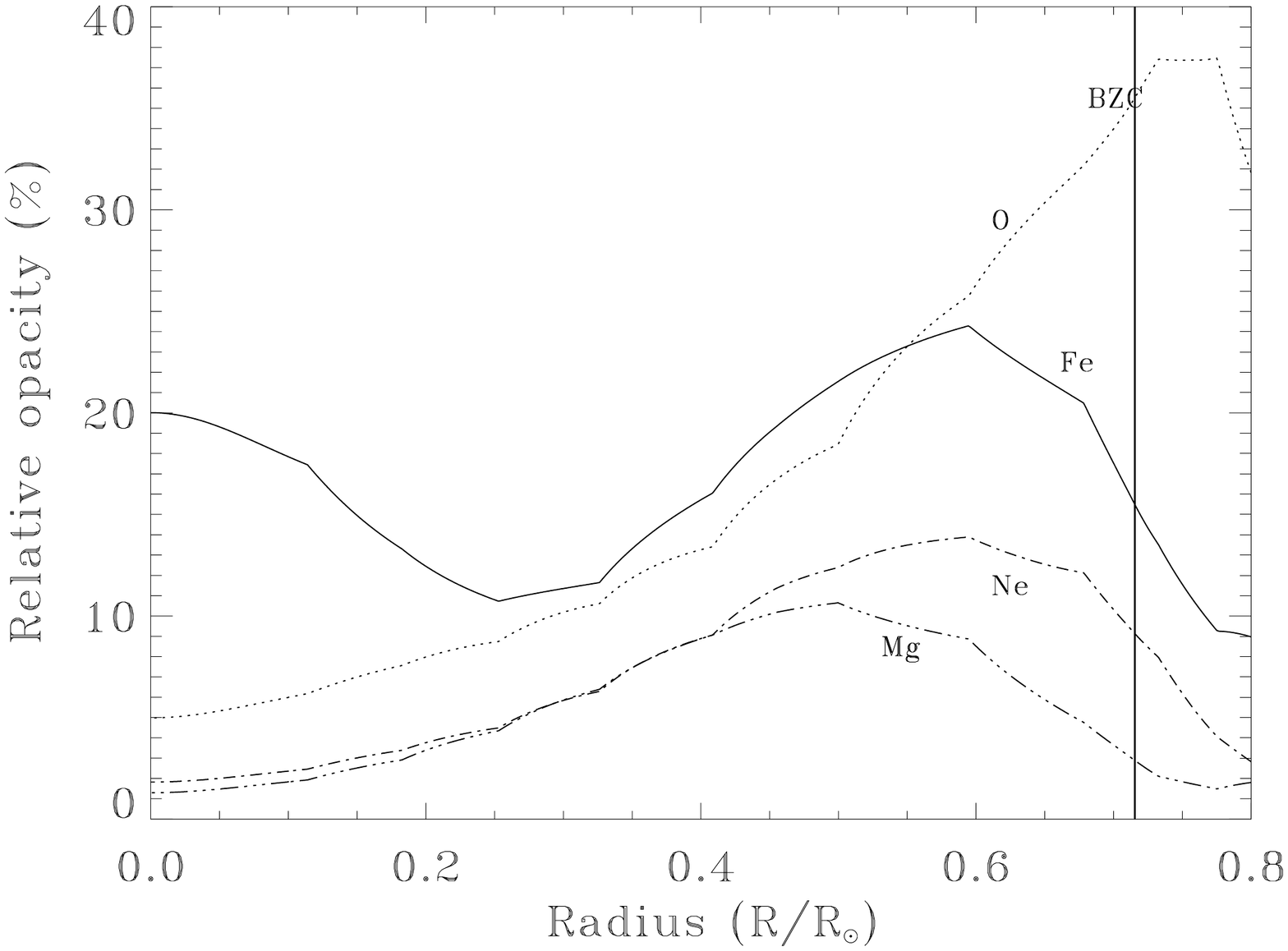}
\vspace{-0.5cm} \hspace{-1cm} \includegraphics[ width=6cm, angle=90]{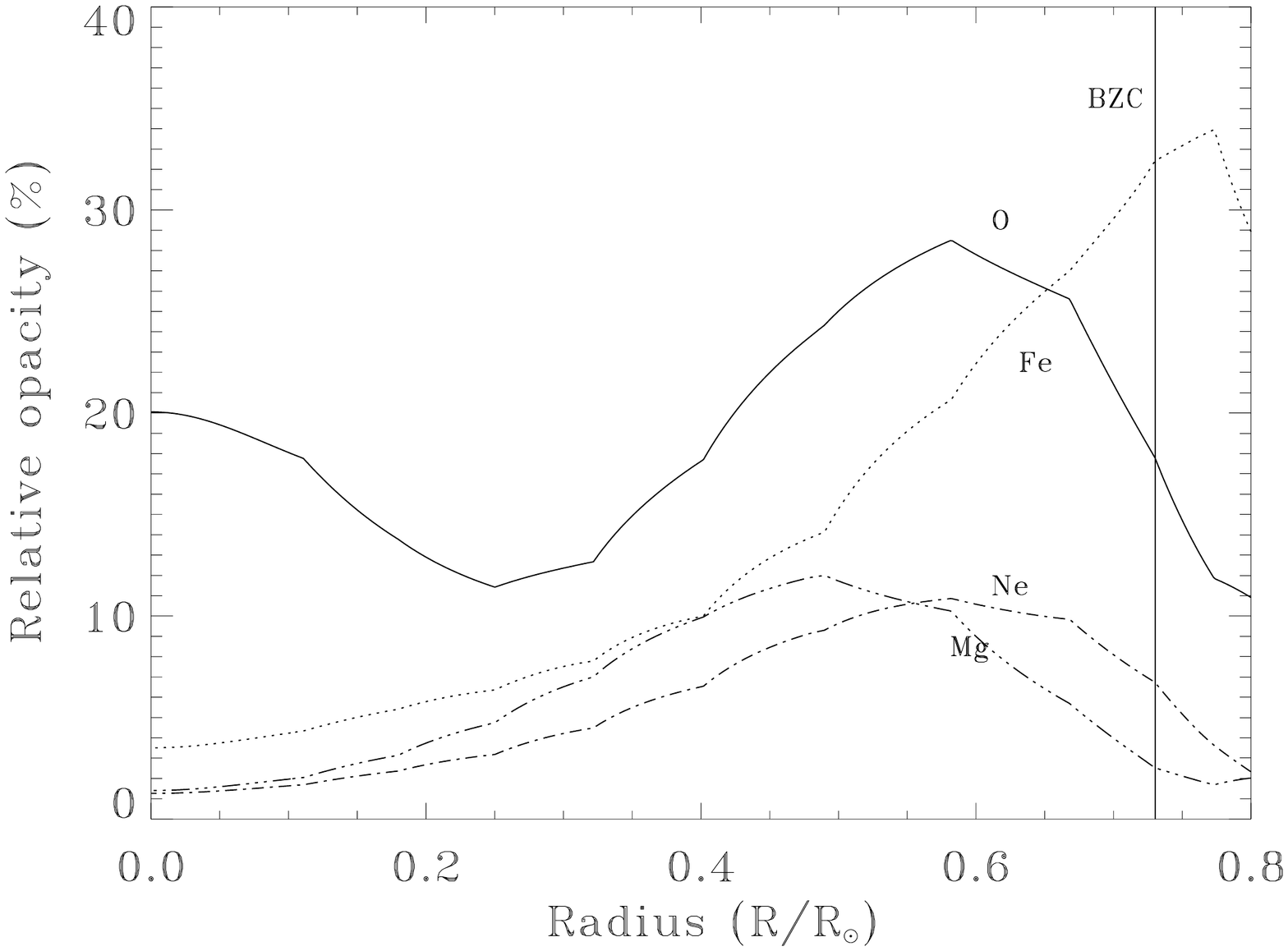}
\vspace{-0.5cm} 
\caption{Relative contribution to the total opacity of the most abundant species which play an important role in the opacity of the radiative zone after  hydrogen and helium. This estimate uses the OPAL opacity calculations for the Grevesse and Asplund composition differing by the content of C, N, O. }
\end{figure}
 
 For higher stellar masses the central temperature increases and all the species are totally ionized (see for example figure 2 of \cite{Courtaud} or 31 of \cite{TC93}). In these stars the core is convective due to the amount of energy produced by the role of CNO, so the lifetime of these stars is mainly driven by the nuclear reaction rates rather than by opacity coefficients. Figure 4 illustrates this point in showing the role of iron (for a solar composition) at the limit of the convective core for different stellar masses.
\begin{figure}
\begin{center}
\includegraphics[width=7cm, angle=90]{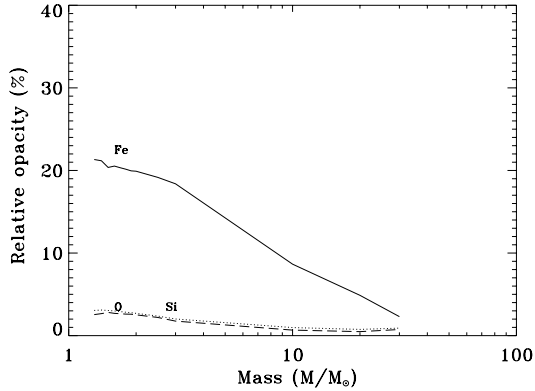} 
\caption{ 
Relative role of iron, oxygen and silicium (for a solar composition) at the limit of the convective core for different stellar masses.}
\end{center}
\end{figure}

\begin{figure}[t]
\begin{center}
\includegraphics[width=6cm]{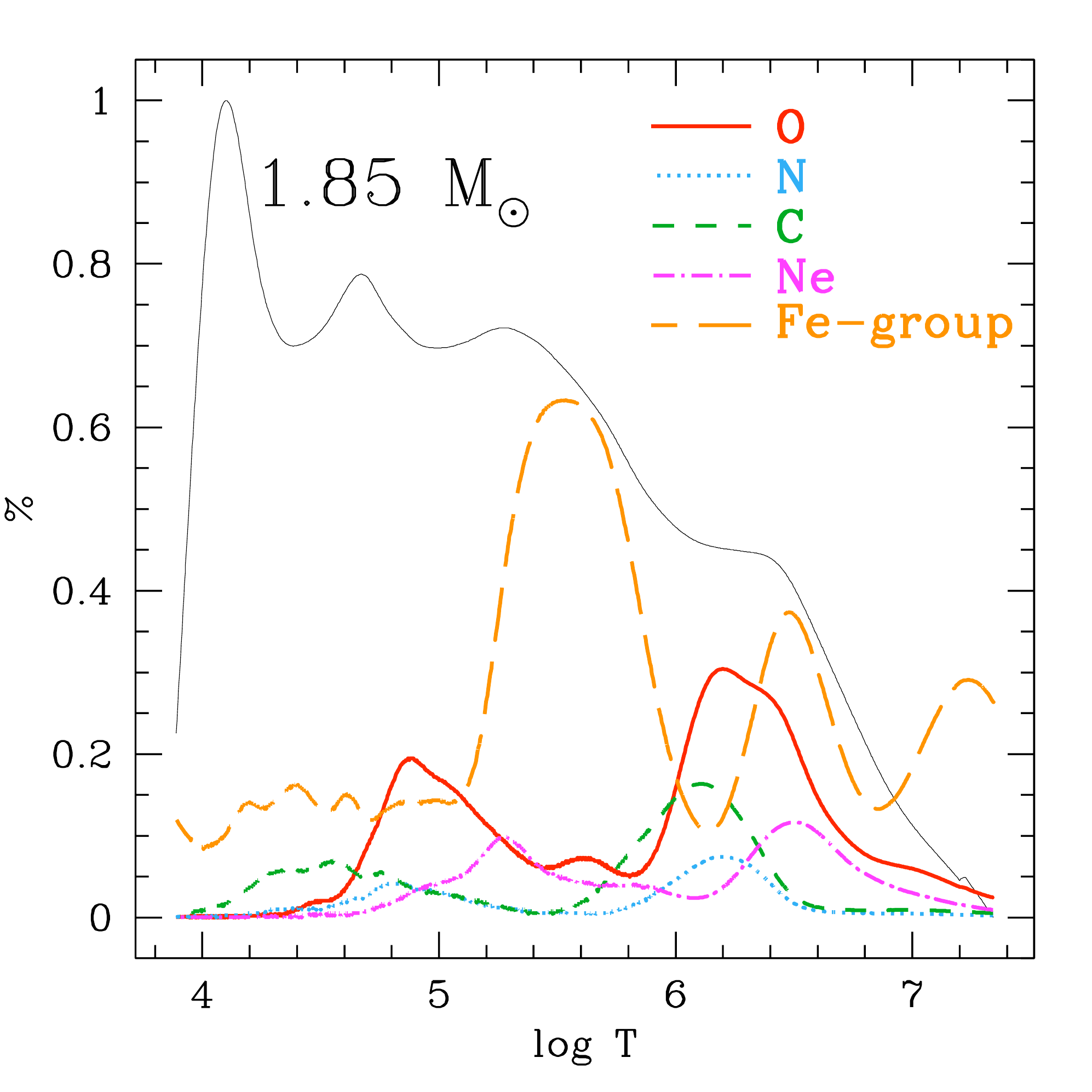}
\includegraphics[width=6cm]{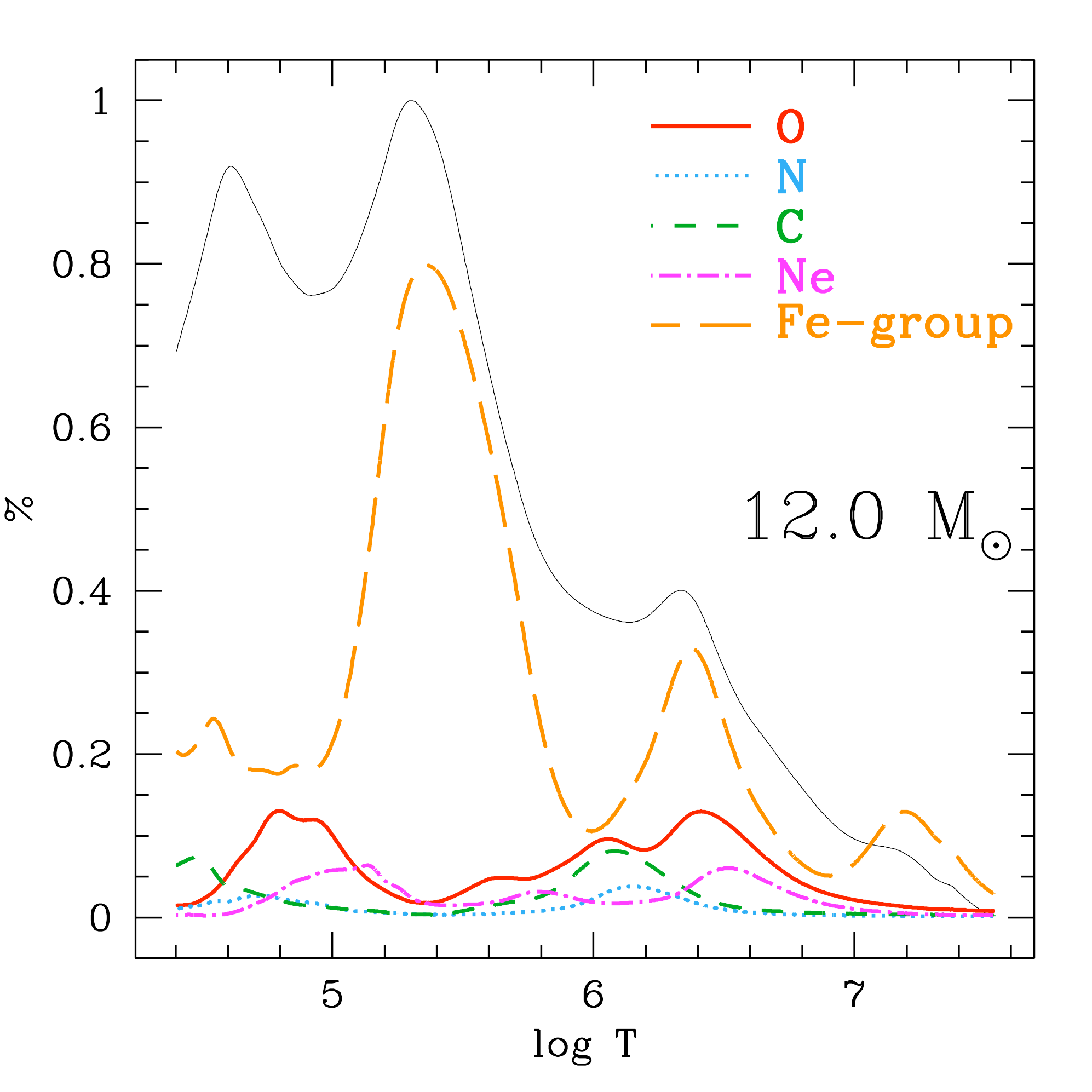}
\caption{  Relative role to the total opacity of the different elements for 1.85 and 12 M$_\odot$ through the whole star. The profile of the total Rosseland opacity of the mixture (in AU) is shown by a thin black solid line. The important role of iron is clearly visible in the external radiative zone and creates the $\kappa$ mechanism when these layers extend or contract with time \cite{Montalban}. (See the color version of the figure in the electronic file).}
\end{center}
\end{figure}
\subsection{The microscopic diffusion of heavy elements and the $\kappa$ mechanism}
We  now examine two other important roles played by the opacity coefficients.
The life of solar-like stars may reach several billions of years. During this long period of time, the different species migrate slightly towards the stellar core due to the gravitational settling of the elements heavier than hydrogen. This effect is extremely slow, resulting for the solar case in 10-15\% depletion at the stellar surface after billion years well measured for helium. In stars a little more massive stars than the Sun, this effect is partly compensated by  the levitation of some species due to radiative pressure  \cite{Michaud, Turcotte}.
To  estimate the respective role of gravity and pressure, one needs to calculate the radiative acceleration for each species k:
$$g_{rad}(k)= {F_R \over c} {M \over M(k)} \kappa_R \gamma (k) \rm \,where \;  \gamma (k)= \int {\kappa_\nu (k) \over \kappa_\nu (total)} f_\nu d \nu \eqno(4) $$
where $F_R$ is the radiative flux. In such a calculation the detailed of the opacity spectrum is needed and $f_\nu$ is a function of the frequency and of the temperature. It seems that there is still some large discrepancies in the radiative acceleration between the two opacity tables available for the international stellar community, the OPAL and the OP tables   \cite{Badnell}. From carbon to iron one may notice some differences up to 50\% between 1.2 eV and 100 eV, in particular for iron  \cite{Delahaye05}, the origin of the difference is not yet understood and could be investigated by some laboratory experiment.

Another important process, called the $\kappa$ mechanism, is at the origin of the  pulsation of intermediate stars.  There is a renewal of interest for that mechanism  since the launch of COROT (2006) which is the first satellite dedicated to the development of asteroseismology   \cite{Michel} and search for exoplanets. This mechanism is responsible for the pulsational instability of  stars between 1.5 to 20 M$_\odot$ and justifies some explanation.
 
In the external optically thick layers, the diffusion approximation is justified and:
$$  \delta \rm div F_R = {1 \over  { 4 \pi r^2}} {d \delta L_r \over dr}  \, where \, {\delta L_r\over L_r} = {{dr \over d Ln T} - {d \kappa \over \kappa} + 4 \large( {\delta T \over T} + {\delta r \over r} \large)} \eqno(5)$$
The first term describes the radiative dissipation and in fact stabilizes the star, the term ${\delta T \over T}$ describes the direct influence of the temperature variation on the luminosity,  this effect is generally small, the last term contributes  to the instability because in case of the compression the radiating area is reduced. The $\kappa$ mechanism will work if the opacity varies more quickly  with radius than the other terms. Two kinds of excitation are possible: the opacity bump is due  to the partial ionization of He ($He^+$ to He$^{++} $) at log T approximatively 4.5 and excites the evolved cepheids, RR Lyrae  (low mass evolved stars) and the $\delta$ Scuti  stars  (1.5-2.5 M$\odot$ main sequence stars) or the opacity bump is due to the M shell of Fe  at log T approximatively 5.2 and excites  the $\beta$ Cepheids stars (7-20 M$\odot$ main sequence stars), SPB (3-9 M$\odot$ main sequence stars)  and sdB stars (helium core, 0.5-1.4 M$\odot$). In a   $\beta$ Cepheid star of 8 M$\odot$ the  total opacity reaches the value of 6.95 $\rm cm^2g^{-1}$  due to the bound-bound process associated to the partial ionization of iron at T= 195800 K (1eV= 11600 K) and for a density of 6.6 $10^{-7} \rm g/cm^3$. We show on Figure 5 the contribution of iron and some other elements in the external radiative zone of different stars from 1.8 to 12 M$_\odot$  \cite{Montalban}.  

The Cepheid stars are known to be used to estimate the stellar distances. They have been better understood after a revision by  a factor 2 of the opacity  for species of Z$>2$ confirmed by the first measurements of absorption coefficients. The  $\beta$ Cepheids stars are now of first interest due to the beautiful new asteroseismic observations and the correlation between the frequencies of the excited modes and the mass of these stars for a given age. 

These stars are now used to estimate the age of several young open clusters  \cite{Balona}. Nevertheless the labeling of the detected frequencies is not an easy task so complementary observations or measurements are needed. The difference between OP and OPAL tables leads to some interesting differences in the stellar surface temperature and it is presently believed in the seismic community that the use of OP opacities seems to help the interpretation of  the observed frequencies \cite{Pamyatnykh}.

\section{From theoretical estimates to laboratory experiments}
We deduce from the above stellar situation two kinds of conditions which will benefit from the present and coming  large laser facilities.

The interest to describe better the physics of the stellar envelope is justified by the study of more and more pulsating stars (COROT, KEPLER) with the objectives to put constraints on the corresponding stellar interiors. This fact leads to study relatively low temperature (about 20 eV) and very low density. Of course these densities are too low to be considered presently and one needs to search equivalent situation at higher density. This kind of investigation could be done on medium size lasers like the LULI2000  in France and also GEKKO XII and VULCAN, an interesting case is shown in \cite{Loisel} when we compare different neighbors of relatively high Z at 20 eV and 3 mg/cm$^3$. For the astrophysical purpose we will enlarge the spectrum domain to XUV between 10 to 100 eV to cover the most important stellar region where the Rosseland mean value is maximal. For these conditions, OP and OPAL spectra present some different behaviors and we are preparing the best experimental conditions to resolve this discrepancy and improve the stellar interpretation of the seismic data \cite{Loisel2}. 

The second case is connected  to opacity coefficients for the internal structure of radiative zones of solar-like stars. The temperature is generally greater than 100 eV and the density greater than fraction of solid density. If we want to investigate different cases, one needs to work on coming larger facilities (LIL +PETAL, OMEGA EP, FIREX II, LMJ, NIF) because the difficulty is mainly to reach the equivalent high density at LTE with a precise diagnostic. In these cases, on needs to use an X-ray spectrometer to measure the frequency band necessary for the astrophysical purpose or to find some equivalence to check the related plasma properties. Let consider this second case in more details.

\subsection{Degree of ionization of the different species in the solar plasma}
The composition of the solar plasma is by definition different than the plasma produced in laboratory. So we estimate here the differences.
Let us first reduce the solar plasma to a 4 components plasma formed by H, He, O and Fe, with the respective  fraction numbers are He = 0.08, O = 5 10$^{-4}$, Fe =  2.8 10$^{^-5}$, hydrogen accounting for the remainder (following approximatively the solar composition).  Clearly in the solar plasma, the free electrons come from hydrogen so it is interesting to calculate the degree of ionization of   Fe and O  in considering pure iron or oxygen or in considering the simplified solar composition. The calculations have been made for two conditions: at the basis of the convective zone where the temperature is about 193 eV and the density 0.2 g/cm$^3$ and for the iron at the maximum of the iron opacity peak around 400 eV and a density of 2.5 g/cm$^3$. 
\begin{table*}
\begin{center}
\caption[]{\label{table 1} Average degree of ionisation for iron and oxygen for different density and temperature and mixture conditions obtained by three calculations:  FLYCHK (F),  OP  (OP) and a mixture model using Thomas Fermi (TF). We consider different cases:  pure iron or oxygen,  the same element in a mixture of H+ Fe (or O) and  the same element in a mixture of 4 elements.}    	
\vspace{0.2cm} 
\begin{tabular}{p{2.cm}*{7}{c}}
\hline
T                    &$N_e $           & $\rm <Z> F $         &   $\rm  <Z> F$   &$<Z> $ OP &  $\rm <Z> \,TF$ \\
$ eV$ &      $cm^{-3 }$   &             $Fe$       &       $ H+Fe$        & Fe      &  $  \small {H,He, O, Fe}$     \\
\hline
172	& $10^{23}$ 	&    15.28            &    15.33       &  15.86               &    16.06\\
193    &  $10^{23}$ 	&    16.32              &    16.38      & 16.73         &    16.88\\
217	 & $10^{23}$ 	&    17.37              &    17.42       &   17.64 &         17.72\\
\hline
380		& $1.2 \,10^{24}$ 	&    19.            &    19.03               &  19.92 (385 eV) &     18.86               \\
400			& 1.2 \,$10^{24}$ 	&    19.38              &    19.42       & -  &   19.18        \\
420			& 1.2 \,$10^{24}$ 	&    19.74              &    19.77       &   20.67 ( 432 eV)      &    19.48           \\
\hline
\hline
T                    &$N_e $           & $\rm <Z> F $         &   $\rm  <Z> F$   &$<Z> $ OP &  $\rm <Z> \,TF$ \\
$ eV$ &      $cm^{-3 }$   &             $O$       &       $ H+O$        & O      &  $  \small {H,He, O, Fe}$     \\

172	& $10^{23}$ 	&    7.            &    7.03       &      7.03       &    7.11\\
193    &  $10^{23}$ 	&    7.28            &    7.30      &  7.29        &    7.24\\
212	 & $10^{23}$ 	&     7.42             &    7.48       &  7.5 &         7.34\\
\hline
\end{tabular}
\end{center}
\end{table*}

Table 1 shows the averaged degree of ionization for iron and oxygen obtained with different codes: the FLYCHK code which provides ionization and population distributions of plasmas using a schematic atomic structures and scaled hydrogenic cross- sections \cite{Chung} that we use for one or two components (H, Fe or H, O), a mixture  model using Thomas Fermi averaged ionization \cite{Gilles07} which allows us to consider the 4 components and also the OP code where we consider only pure Fe and O. One notices that the choice of the mixture does not  significantly change the degree of ionization when T and N$_e$ (electron number) are fixed but that this degree of ionization depends on the hypotheses of the model used (mainly for iron). The density of the corresponding pure iron measurement at 193 eV for the same averaged degree of ionization can be obtained in replacing the density of the plasma  by  $\rho_H {{56} \over  { \xi^*}}$ where $\xi*$ is the ionization degree. In the discussed case the density of the pure iron must be = 0.6 g/cm$^3$  instead of the value of 0.2 g/cm$^3$ of the solar plasma. Of course an opacity experiment of a mixture of elements could be also another interesting experiment to perform.

\subsection{Sensitivity of the absorption spectra to the temperature and density conditions}
The  Z pinch facility of Sandia National Laboratory has been used for the first comparison between experiment and  calculations \cite{Bailey}. The experiment has been done at T$_e$= 156 $\pm$ 6 eV and N$_e$ = 6.9 $\pm$ 1.7 10$^{21}$ cm$^{-3}$. These thermodynamic conditions seem not so  far from the solar conditions discussed in table 2 even it is always difficult to determine the real conditions of measurements \cite{Cheng}. 

\begin{figure}[t]
\begin{center}
\includegraphics[width=14cm]{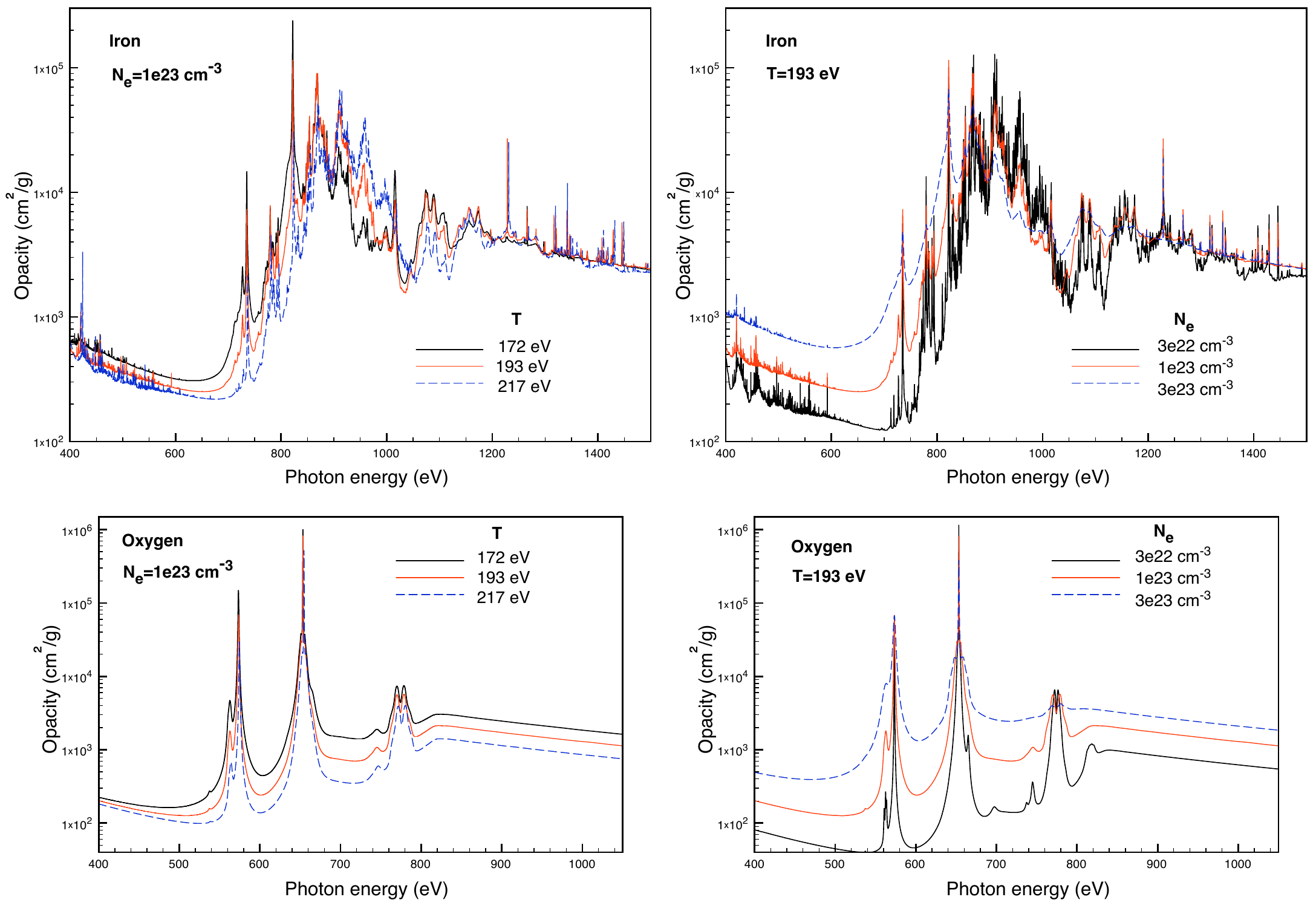} 
\vspace{-0.5cm} 
\caption{Comparison of the individual iron and oxygen spectra for different conditions of temperature and density around the BCZ solar conditions in using the OP tables. (See the color version of the figure in the electronic version).}
\end{center}
\end{figure}
The experimental difficulties in getting a detailed comparison with  the spectra used in astrophysics are numerous. First one needs to determine without ambiguity the proper temperature and density of the experiment, then one needs to obtain the proper  level of transmission to look to the absolute value of the detailed spectra.
We illustrate on Figure 5 how the iron and oxygen spectra evolve for the OP calculations with temperature and density in a range probably larger than the uncertainty we may have for these two values.  Doing so we illustrate directly the complexity of the iron spectra in comparison with the oxygen spectra. Of course the interesting spectrum range is around 800 eV which corresponds to the maximum of the mean Rosseland value.  Although dominant peaks present small variations, peak to valley ratio is clearly different. They are sensitive to temperature variation of 10\% and yields to a variation of 6\% in Rosseland opacity in the case of iron,  the spectra show clearly the difference in  ionization numbers. The effect on the Rosseland mean value for a change of density by a factor 3 is between 8 and 13\%.   

So  several measurements of  absorption spectra obtained on different facilities will be useful. They require a good determination of the plasma temperature and density which can be delivered by petawatt lasers  and a  well calibrated transmission. These experiments will help to disentangle the origins of the observed  differences between the calculations from OP and OPAL for some  astrophysical conditions. They will also   give strong confidence in these very complex atomic calculations and will contribute to build a more sophisticated view of stars which must include microscopic and macroscopic phenomena to reproduce all the seismic observables: sound speed, superficial helium content, internal rotation, circulations, magnetic field.


\begin{thebibliography}{00}
  \bibitem{TC93}Turck-Chi\`eze, S., D\"appen, W., Fossat, E., Provost, J.,
Schatzman, E. \& Vignaud, D., Physics Rep. 230 (1993) 57.
 \bibitem{Vorontsov02}Vorontsov, S.V., Christensen-Dalsgaard, J, Schou, J. et al., Science 296 (2002) 101
 \bibitem{Mathur08}  Mathur, S. et al.,  A\&A 484 (2008) 517\\
 \bibitem{Townsend07}  Townsend, R. et al.,  MNRAS 382 (2007) 139\\
\bibitem{Ud-Doula08}  Ud-Doula, A. et al.,  MNRAS 385 (2008) 97\\  
\bibitem{CDB91} Christensen-Dalsgaard, J. \& Berthomieu, G.,  in Solar interior and atmosphere (2001) eds A. N. Cox, 
  W.C. Livingston and M. Matthews, Space Science Series, p 401.
\bibitem{TC01a}Turck-Chi\`eze, S. et al., ApJ 555 (2001a) L69
\bibitem{TC01b} Turck-Chi\`eze, S., Nghiem, P., Couvidat, S., Turcotte, S., Sol. Phys. 200 (2001) 323
\bibitem{TC01}Turck-Chi\`eze, S., Nucl. Phys. B (Proc. Suppl.) 91 (2001) 73
\bibitem{TC04}Turck-Chi\`eze, S. et al., Phys. Rev. Lett 93 (~2004) 211102
\bibitem{Delahaye} Delahaye, F. \& Pinsonneault, M., ApJ  649 (2006) 529
\bibitem{Turck08} Turck-Chi\`eze, S., Nghiem, P. P. A. \& Mathis, S., JPh: CS 118  (2008) 12030
\bibitem{Courtaud} Courtaud, D. et al., Sol. Phys. 128 (1990)  49
\bibitem{Asplund} Asplund, M. et al.,  A\&A 417 (2004) 751
\bibitem{Iglesias} Iglesias, C.A.  \& Rogers, F. J.,  ApJ  443 (1995) 469
\bibitem{Cox} Cox, A. N.  \& Tabor, J. E.,  Astron. J. Suppl. Ser. 31 (1976) 271
\bibitem{Rogers}  Rogers, F. J. \& Iglesias, C.A.,  ApJS. 79 (1992) 507
\bibitem{TC05} Turck-Chi\`eze, S., Couvidat, S. \& Piau, L., EAS Publications series 17 (2005) 149
\bibitem{TC08} Turck-Chi\`eze, S. \& Talon, S., Adv. Space. Rev. 41 (2008) 855
\bibitem{Seaton}  Seaton M. J. \& Badnell, N. R.,  MNRAS 354 (2004)  457
\bibitem{Badnell}  Badnell, N. R., et al.,  MNRAS 360 (2005) 458
\bibitem{Delahaye05}  Delahaye, F. \& Pinsonneault, M.,  ApJ 625 (2005) 533
\bibitem{Michaud} Proffitt, C.R.  \& Michaud, G. ApJ 371 (1991) 584
\bibitem{Turcotte} Turcotte, S. et al.,  ApJ 504 (1998) 559
\bibitem{Montalban} Montalban, J. \& Miglio, A., Comm. in Asteroseimology, Proceeding of the Wroclaw, Helas workshop, 157 (2009) 160
 \bibitem{Michel} Michel, E. et al., Comm. Astr. 157 (2008) 69
 \bibitem{Balona} Balona, L. A., Dziembowski, W. A. \& Pamyatnykh, A.A.,  MNRAS 289 (1997) 25
  \bibitem{Pamyatnykh} Pamyatnykh A.A., Acta Astronomica 49 (1999) 119
 \bibitem{Loisel} Loisel, G. et al.,  HEPD (2009) 5, 173
 \bibitem{Loisel2} Loisel, G., Turck-Chi\`eze, S., Delahaye, F. \& Piau,  ApJ (2009) to be published
 \bibitem{Chung} Chung, H. K. et al.,  HEDP 1 (2005) 3
 \bibitem{Gilles07} Gilles, D. et al., HEDP 3 (2007) 95 
 \bibitem{Bailey} Bailey, J. et al., Phys. Rev. Lett. 99 (2007) 5002
  \bibitem{Cheng} Cheng, G. and Jiaolong, Z., Phys. Rev. E 78 (2008) 046407
 

 
 


\end{thebibliography}
\end{document}